\documentclass{aastex}
\usepackage{emulateapj5}

\newcommand{\NHA}{log$([$NII$]\lambda6583/$H$\alpha)\;$}
\newcommand{\nha}{$[$NII$]\lambda6583/$H$\alpha\;$}
\newcommand{\OHB}{log$([$OIII$]\lambda5007/$H$\beta)\;$}
\newcommand{\ohb}{$[$OIII$]\lambda5007/$H$\beta\;$}

\submitted{To appear in the May, 2002 AJ}

\lefthead{Melbourne \& Salzer}
\righthead{Metal Abundances of KISS Galaxies I.}

\begin{document}

\title{Metal Abundances of KISS Galaxies I.\\
Coarse Metal Abundances and the Metallicity-Luminosity Relation.}

\author{Jason Melbourne\altaffilmark{1} and John J. Salzer}
\affil{Astronomy Department, Wesleyan University, Middletown, CT
  06459}
\altaffiltext{1}{Present address: Department of Astronomy and Astrophysics,
  University of  California, Santa Cruz, Santa Cruz, CA 95064}  
\email{jmel@ucolick.org; slaz@astro.wesleyan.edu}

\begin{abstract}
We derive metal abundance estimates for a large sample of starbursting
emission-line galaxies (ELGs).  Our sample is drawn from the KPNO
International Spectroscopic Survey (KISS) which has discovered over 2000
ELG candidates to date.  Follow-up optical spectra have
been obtained for 
$\sim900$ of these objects.  A three step process is used to obtain
metal abundances for these galaxies.  We first calculate accurate
nebular abundances for 12 galaxies whose 
spectra  cover the full optical region  from [OII]$\lambda\lambda$3726,29
to beyond [SII]$\lambda\lambda$6717,31 and include detection of 
[OIII]$\lambda$4363.  Using secondary metallicity indicators
$R_{23}$ and $p_3$, we calculate metallicities for an additional 
59 galaxies with spectra that cover a similar wavelength range but lack
[OIII]$\lambda$4363.  The results are used to calibrate
relations between metallicity and two readily observed emission-line
ratios, which allow us to  estimate coarse
metallicities for 519 galaxies in total.  The uncertainty in
these latter abundance estimates is 0.16 dex.  From the large,
homogeneously observed sample of star-forming galaxies we
identify low metallicity candidates for future study and investigate the 
metallicity-luminosity relation.  We find a linear metallicity-luminosity
relation of the following form: $ 12 + $log(O/H)$ =  4.059  - 0.240 M_B,$ 
with an RMS scatter of $0.252$.  This result implies that the slope of
the metallicity-luminosity relation is steeper than when dwarf
galaxies are considered alone, and may be evidence that the
relationship is not linear over the full
luminosity range of the sample. 
\end{abstract}


\section{Introduction}
Observing metal abundances in galaxies is a valuable probe of galaxian
star-formation histories and chemical evolution.  
Assuming that all galaxies begin with the 
same primordial abundances of elements, roughly 75\% hydrogen and 25\%
helium,  measurements of heavier
elements indicate subsequent star formation accompanied by  supernova
explosions which  enrich their surroundings with metals.
Recent studies show that metallicity is linked with
galaxy luminosity (Skillman et al. 1989; Zaritsky et al. 1994; Richer
and McCall 1995; Garnett et al. 1997; Hunter and Hoffman 1999;
Pilyugin and Ferrini 2000; Pilyugin 2001b) in that more luminous
galaxies tend to 
be more metal rich than less luminous galaxies.  This may indicate an 
evolutionary trend with several possible explanations. For example, a simple
closed box model gives rise to such a  trend (Hidalgo-Gamez and
Olofsson 1998).  Alternatively, evidence for large disruptions in the
gas content of 
galaxies from  supernova explosions may play a significant role.
Specifically, supernovae have been proposed as a mechanism for removing
large amounts of metal-enriched gas from low-mass systems (e.g., Mac
Low and Ferrara 1999).  This in turn may 
explain the relative dearth of metals in dwarf galaxies.  In addition,
higher astration levels in more 
luminous galaxies may contribute to their increased metallicity per
mass (Pilyugin and Ferrini 2000). 

The KPNO International Spectroscopic Survey (KISS) has identified
over 2000 emission-line galaxy (ELG) candidates ranging 
in absolute magnitude from M$_B$ = -22 to -12 (Salzer et al. 2000, 2001).
As such, it provides a 
large sample of galaxies for which metallicities can be
derived and from which the metallicity-luminosity relation can be
studied. The survey lists include massive starburst-nucleus galaxies,  
intermediate-mass irregular galaxies, low-mass dwarf irregulars and 
blue compact dwarfs.  In this work we derive coarse metallicity
estimates  for the 519 starburst galaxies which have adequate quality
follow-up 
spectra. The results provide a list of low metallicity candidates for
future study as well as the largest sample of galaxies to date for use
in studying the metallicity-luminosity relation  

Most previous studies of the metallicity-luminosity relation have
concentrated on low metallicity galaxies where accurate abundances are
readily available.  Studies of irregulars with current star formation
(Skillman et al. 1989; Richer and McCall 1995)
show evidence for a linear relation.  However, another recent study of dwarf
irregulars  (Hidalgo-Gamez and Olofsson 1998) does not support these
results. In re-examining the Hidalgo-Gamez and Olofsson data, Pilyugin (2001b)
believes that noise in the [OIII]$\lambda$4363 line is responsible for
the lack of a relation in their sample.

On the high mass end, Garnett et al. (1997) 
compiled a data set of 29 luminous spirals, with metallicities derived
from HII regions.  The data set is taken from
Vila-Costas and Edmunds (1992), Zaritsky et al. (1994) and Ryder (1995).
They find that these more massive galaxies 
follow a similar trend to dwarf irregulars of increasing
metallicity with luminosity.  Garnett et al. do not offer a specific fit
to the relation, but rather state that the high luminosity result maps
smoothly onto the dwarf irregular relations such as that of Skillman
et al.  The same conclusions are drawn by Zaritsky et al. (1994).
Pilyugin and Ferrini (2000) combine the low and high mass galaxy
samples to provide a new fit to the metallicity-luminosity relation
and show a somewhat steeper relation than Skillman et al (1989).  

By using the large KISS sample of galaxies we investigate whether the 
metallicity-luminosity relation is actually present on all mass,
luminosity and metallicity scales.  Our abundance
determinations are coarse and we use all morphological types for our
metallicity-luminosity relation.  Therefore our relation 
contains significantly more scatter than previous results.  However,
our luminosity range goes three magnitudes brighter 
than most previous studies, and we use up to 20 times more galaxies, giving
us a better view of the overall form of the metallicity-luminosity
relation, as well as its intrinsic scatter.  We confirm the existence
of a metallicity-luminosity relation that varies 
smoothly from massive luminous galaxies, M$_B < -21$, to blue compact dwarfs,
M$_B > -16$.  However,
when the metallicity-luminosity relationship is calculated using the 
wide range of galaxy types found in the 
KISS sample, the slope of the relation is steeper than indicated by the 
studies of dwarf galaxies such as Skillman  et al. and Richer and McCall.
As a result we find that a simple extrapolation of the dwarf galaxy
relationship to 
more massive systems is inappropriate.  This may be an indication 
that the overall form of the metallicity-luminosity relation is not
linear but rather requires a higher order polynomial to fit the data.  
It may also indicate that different galaxy types obey 
different metallicity-luminosity relations.

\section{Observations and Data Reduction} 

\subsection{Observations}
Drawing from the KISS sample of 
emission-line galaxies, we gathered  
both imaging and spectral data for $\sim$900 galaxies over a four
year period from 1998 - 2001. Photometry is presented in the survey lists
(Salzer et al. 2001; 2002a; Gronwall et al. 2002a) while
results of spectral 
follow-up are given in a series of papers (Melbourne et al. 2002,
hereafter Paper II; Salzer et al. 2002b; Wegner et al. 2002). A
summary of the overall spectroscopic properties of the KISS ELGs will
be presented in Gronwall et al. (2002b).  

\begin{figure*}[htp]
\vskip -1.7in
\epsfxsize=6.5in
\hskip 0.5in
\epsffile{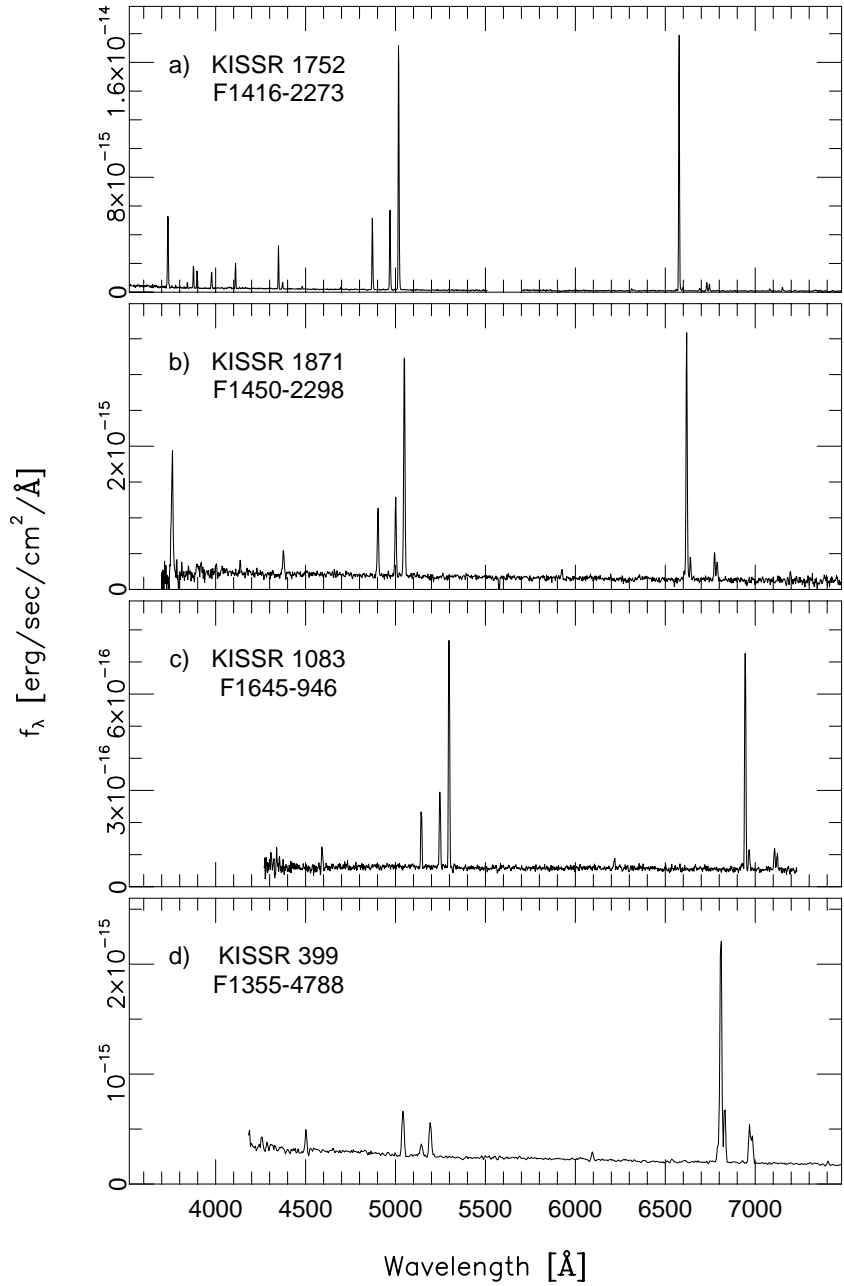}
\vskip -0.7in
\figcaption[jmelbourne.fig1.eps]{Sample spectra from the KISS
  archive:  a) A Group I spectrum taken at the Lick 3m
  telescope.  It covers the full optical region from [OII]$\lambda\lambda$3726,29
to beyond [SII]$\lambda\lambda$6717,31 and includes detection of the
[OIII]$\lambda$4363 line necessary for accurate metal abundances.
 b) A Group II spectrum taken at the Kitt Peak 2.1m
telescope.  It also covers the full optical region but does not have
the signal-to-noise ratio necessary for the detection of
[OIII]$\lambda$4363. c) A Group III spectrum taken
with the HET 9m telescope.  It has a high signal-to-noise ratio because of
the large telescope aperture, but does not cover the full optical region and is
specifically lacking [OII]$\lambda\lambda$3726,29.  d) Also a group
III spectrum, taken with the MDM 2.4m. It has lower dispersion;
the nitrogen lines are blended with H$\alpha$, and the
sulfur doublet is also blended.  \label{fig:spectra}}
\end{figure*}

The spectral data can be classified
into 3 groups based on spectral coverage and quality. 
Group I spectra, an example of which is given in Figure
\ref{fig:spectra}a, cover the full optical region from
[OII]$\lambda\lambda$3726,29 
to beyond [SII]$\lambda\lambda$6717,31 and contain the 
[OIII]$\lambda$4363 line necessary for accurate abundance
measurements.   Abundances are calculated for Group I
galaxies using the [OIII]$\lambda$4363 line as an electron temperature
indicator.  All 12 objects in this group were observed with the Lick 3m
telescope.  The spectroscopic data and abundance analysis are
presented in detail in Paper II.   

Group II data also cover the full optical region from
[OII]$\lambda\lambda$3726,29 
to beyond [SII]$\lambda\lambda$6717,31, but do not necessarily contain the 
[OIII]$\lambda$4363 line needed for accurate abundance
measurements.  Metal abundance estimates are derived from
the strong oxygen lines [OII]$\lambda\lambda$3726,29
and [OIII]$\lambda\lambda$4959,5007 using the secondary metallicity
indicators $R_{23}$ (Pagel et al. 1979) and $p_3$ (Pilyugin 2000).
The data were obtained at 
the Lick 3m, the APO 3.5m and KPNO 2.1m telescopes and include 59 additional
objects.  A sample spectrum is shown in Figure \ref{fig:spectra}b.

Group III spectra do not reach blueward to the
[OII]$\lambda\lambda$3726,29 line.  
Metal abundances are derived for these galaxies based on
empirical relations between the \ohb line ratio and metallicity and
the \nha line ratio
and metallicity.  Data in this category were obtained from the
above mentioned telescopes as well as the WIYN 3.5m, the MDM 2.4m and the
Hobbey-Eberly 9m (HET). 
As the HET is a 9m-class telescope, the spectra
have a high signal-to-noise ratio as shown in Figure
\ref{fig:spectra}c. Unfortunately, they do not extend much bluer than
H$\gamma$.  The [SII] doublet on the red side is also often beyond the
wavelength range of the spectrograph. The MDM images have a low
dispersion.  Therefore, as is shown in Figure  
\ref{fig:spectra}d, the
[NII]$\lambda$6583 line blends with H$\alpha$, and the
[SII]$\lambda\lambda$6717,31 doublet is also blended. 

Details of the observations, data reduction methods and line measurements
are given in a series of papers which present the results of our
extensive spectroscopic follow-up (Paper II;
Salzer et al. 2002b; Wegner et al. 2002). 

\section{Metallicities of KISS Galaxies}

The most accurate calculation of metallicity in
nebular star-forming regions  uses the
[OIII]$\lambda$4363 line to  measure the electron gas temperature
(Osterbrock 1989; Izotov et al. 1994).  This
method, referred to here as the T$_e$ method, works well for low metallicity
systems (generally 12 + log(O/H) $<$ 8.2) where the
[OIII]$\lambda$4363 line is observable. For systems of higher
metallicity,   [OIII]$\lambda$4363 is often too weak to observe or too
noisy to be trusted.  However,
the strong nebular lines alone contain the necessary information to
arrive at relatively good estimates of the  oxygen abundances in
star-forming regions (Pagel et al. 1979; McGaugh 1991;  
Pilyugin 2000).  The traditional $R_{23}$ method (Edmunds and Pagel 1984) 
typically results in metallicities within 0.2 dex of the more exact
$T_e$ abundances.  Pilyugin (2000) 
improves on the standard $R_{23}$ method by introducing his $p_3$ factor 
calculated from the strong oxygen lines.  The $p_3$ factor replaces the 
temperature as a descriptor of the conditions in the nebula and allows 
his method to correlate with the $T_e$ method to within 0.1 dex for
starbursts with metallicities below 12 + log(O/H) $<$ 7.9.
In this section we make use of the T$_e$, $R_{23}$ and $p_3$ methods
to estimate metallicities  
for 71 KISS galaxies.  These data are then used to show that both the \ohb 
and \nha line ratios correlate with metal  abundance.
We use the relationships to estimate coarse metallicities for the large, 
homogeneously observed KISS sample of galaxies.  The details of
each step follow.

\begin{figure*}[htp]
\vskip -1.7in
\epsfxsize=5.5in
\hskip 1.0in
\epsffile{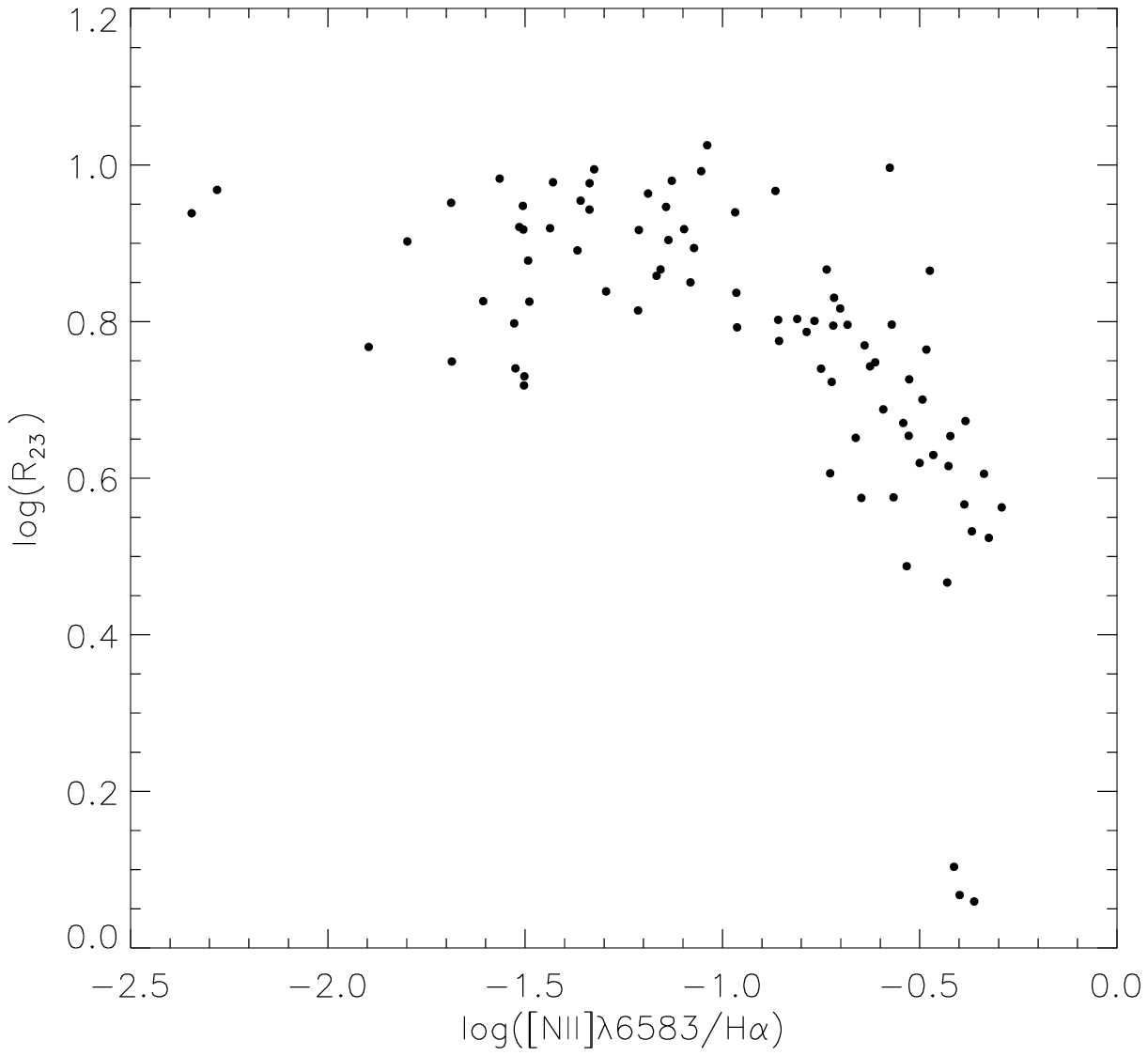}
\vskip -0.7in
\figcaption[jmelbourne.fig2.ps]{This line diagnostic diagram shows the
  relationship between $R_{23}$ and the \nha line ratio.  Metallicity
  varies smoothly along the observed distribution, with high
  metallicity galaxies to the lower right and low metallicity
  galaxies to the upper left.  When \NHA $<$ -1.3 the galaxy is on
  the low metallicity
  branch of the $R_{23}$ parameter. When - 1.3 $<$ \NHA $<$ -1.0 the
  galaxy is in the   turn around region, and when \NHA $>$ -1.0 the
  galaxy is the high metallicity
  branch of the $R_{23}$ parameter. \label{fig:X23NHA}}
\end{figure*}

\subsection{T$_e$ Metallicities}
Spectra of 12 KISS galaxies contain the necessary information for high quality
metal abundance determinations.  Following the standard procedure
(Osterbrock 1989; Izotov et al. 1994) we
calculate the electron density from the [SII] line ratio, and the electron
temperature from the [OIII] line ratio.   Metal abundances are derived
using the IRAF NEBULAR package (de Robertis et al. 1987; Shaw et
al. 1995). Details of the observations, data reductions and analysis
are given in Paper II.  The results  are presented  in Table \ref{table:Te}.
We use the T$_e$ abundance results in Section 3.3 when we
correlate metallicity with emission-line ratios.  

\subsection{Secondary Metallicity Indicators $R_{23}$ and $p_3$ }
The strong oxygen lines
[OIII]$\lambda\lambda$4959,5007 and [OII]$\lambda\lambda$3726,29 contain 
the necessary information to predict the metallicity of an HII region
(Pagel et al. 1979; McGaugh 1991).  Traditionally this has been done by
invoking the $R_{23}$ parameter (Pagel et al. 1979), where;
\begin{equation}
R_{23} = \frac{f([\mbox{OIII}]\lambda4959+\lambda5007)  +
  f([\mbox{OII}]\lambda3726+\lambda3729)}{f(\mbox{H}\beta)}.
\end{equation}
The $R_{23}$ parameter has been correlated with metallicity by measuring oxygen
abundances \emph{via} the T$_e$ method for large samples of galaxies
and HII regions.
A complication with this method is that the dependence of metallicity 
on $R_{23}$ is double valued.  On the low metallicity end, 
12 + log(O/H) $<$ 7.9,  $R_{23}$ 
increases with metallicity.  As the amount of 
oxygen in the nebula increases the strength of the emission leaving the nebula
in the forbidden oxygen lines also increases.  For higher 
metallicities, 12 + log(O/H) $>$ 8.1, $R_{23}$ decreases with 
increasing metallicity.  In this case the 
majority of the energy leaves the nebula in other 
nebular lines.  Between the high and low metallicity branches is the
``turn-around  
region'' where $R_{23}$ ceases to be a good predictor of metallicity.
In the ``turn-around region'' galaxies with the same $R_{23}$ ratio
can have a fairly wide range of metallicities. 

We distinguish between the upper and lower branches of the $R_{23}$
relation by observing the 
\nha line ratio.  This is demonstrated in Figure \ref{fig:X23NHA},  a
line diagnostic diagram plotting  log($R_{23}$) vs. \NHA \hspace{-2mm}.
In this plot, metallicity varies smoothly across the figure with low
abundance galaxies
found in the upper left hand corner and high abundance galaxies
in the lower right.  We adopt the following criteria for
determining on which  
branch of the $R_{23}$ relation a given spectrum lies.  We assign
objects with   
\NHA $< -1.3$ to the low metallicity  branch of the $R_{23}$ relation.  
Objects with \NHA $> -1.0$  are associated with the high 
metallicity branch. Objects 
with -1.3 $<$ \NHA  $<$ -1.0 are considered turn around region
objects.  We do not calculate $R_{23}$ metallicities for these galaxies. 

Unfortunately $R_{23}$ does not correlate perfectly with metallicity.
Galaxies with a given value for $R_{23}$ can  exhibit a range of
T$_e$-method-determined metallicities.  Other
factors have been found to account for the variations.  The spread in
metallicity, especially on the low 
metallicity end, for a given $R_{23}$ value is due 
to both to uncertainty/scatter in the T$_e$ calibration and to
additional parameters such as the initial mass function (IMF) of
the starburst and density variations in the nebula (McGaugh 1991).
The spread in metallicities for a given value of $R_{23}$ at the high
metallicity end is due primarily to a lack of good T$_e$ abundance
measurements.

\begin{figure*}[htp]
\vskip -1.8in
\epsfxsize=5.5in
\hskip 1.0in
\epsffile{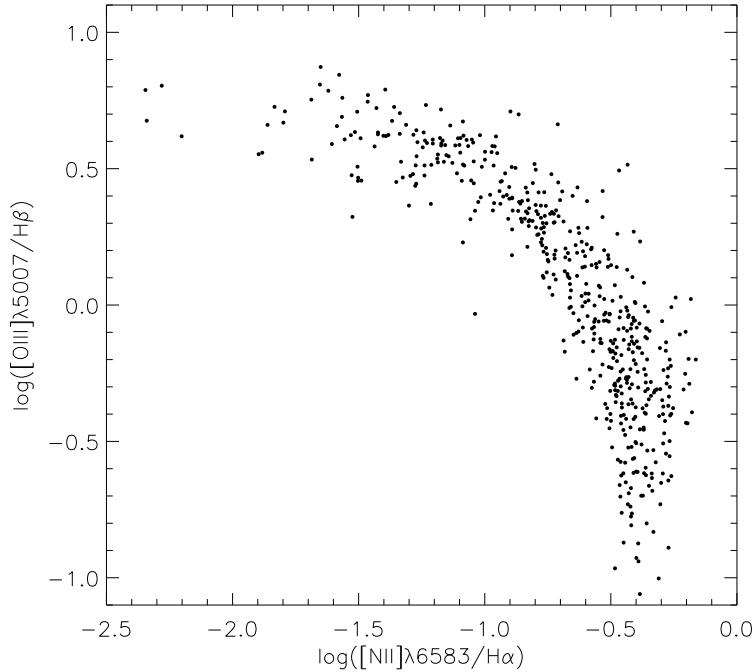}
\vskip -0.7in
 \figcaption[jmelbourne.fig3.ps]{This line diagnostic diagram shows the
  relationship between the \nha line ratio and the \ohb line ratio.
  High metallicity galaxies are found on the lower right where the
  oxygen line ratio is a good predictor of metallicity.  Low
  metallicity galaxies are found to the upper left where the nitrogen
  line ratio is a good metallicity indicator.  
  \label{fig:O3N2}}
\end{figure*}

A recent paper by Pilyugin (2000) demonstrates a method to remove $R_{23}$'s
density and IMF dependencies at the low metallicity end.  Pilyugin
defines a new parameter, $p_3 = X_3 - X_{23}$, where $X_3 =
$log(f([OIII]$\lambda4959+\lambda5007)/f($H$\beta))$ and  $X_{23} =
$log$(R_{23})$. If 
the measure of $X_{23}$ is constant for HII regions of similar oxygen
abundance, then a plot of $p_3$ vs. $X_3$ should give a line with a
slope of 1 for objects of the same metallicity.  When Pilyugin
plotted several HII regions in this way, he found the slope is not 1, 
implying that
$X_{23}$ can vary for objects with the same oxygen abundance.  He 
defined a parameter, $X_3^*$, which is equal to the value of $X_3$
when the data in the $X_3$ vs. $p_3$ plot are extrapolated to
$p_3=0$.  Doing so he found the following relation:
\begin{equation}
X_3^* = X_3 - 2.20 p_3.
\end{equation}
For any observed value of $X_3$ and $p_3$ one can calculate
$X_3^*$.  Pilyugin goes on to demonstrate a correlation between $X_3^*$
and metallicity which agrees with the T$_e$ method to within 0.1 dex.
The $p_3$ method 
effectively removes the systematic uncertainties in the $R_{23}$
method on the low metallicity end.  As he describes it, the temperature
measurement in the T$_e$ method accounts for the state of the 
the IMF and the geometry factors in the nebula.  In this new
method, $p_3$ replaces the temperature as a descriptor of these
influences.  The correlation Pilyugin found which we will 
adopt for objects on the lower metallicity branch of the $R_{23}$
relation is given in by the
following equation:
\begin{equation}
\label{pil}
12 + \mbox{log(O/H)} = 6.35 + 1.45 X_3^*.
\end{equation}

Pilyugin (2001a) derives a similar relation for the upper
metallicity branch.  Unfortunately it is of use only in a limited
metallicity range ($\sim$8.1 to $\sim$8.6 dex).  In addition, fits to
the upper branch  
suffer more from a lack of
good calibration points than from a multi-valued $R_{23}$ relation. 
For simplicity we will adopt the Edmunds and Pagel (1984) fit to the upper
branch.  This fit, as quoted in Pilyugin (2000), can be expressed as follows:
\begin{equation}
12 + \mbox{log(O/H)} = 9.57 - 1.38 \mbox{log}(R_{23}).
\end{equation}

With the two methods, $R_{23}$ and $p_3$, we determine metallicities for 
galaxies which possess spectra that 
contain both the strong [OIII] lines $\lambda\lambda$4959,5007 and the strong 
[OII] line $\lambda\lambda$3726,29.  When \NHA $< -1.3$, we use $p_3$
to derive metallicity.  When  \NHA $> -1.0$ we  use $R_{23}$ to derive
metallicity.  For galaxies with -1.3 $<$ \NHA  $<$ -1.0 we can not use
either method as the object is in the turn-around region. 

As an additional consideration, small uncertainties in
the reddening coefficient, $c_{H\beta}$, can translate to large
errors in the 
[OII]$\lambda3726+\lambda3729$/H$\beta$  line ratio.  Therefore
metallicities derived with the strong-line 
method can be influenced by weak or noisy  H$\beta$ lines. For
example, if we plot
log([OIII]$\lambda4959+\lambda5007$/[OII]$\lambda3726+\lambda3729$)
vs. log($R_{23}$) (McGaugh 1991) for our sample of galaxies,  we find
several objects in unphysical locations of the diagram.  When we plot
the same diagram restricting our sample of 
objects to those with an equivalent width of H$\beta$ greater than 8
\AA, the discrepant points are removed.  Therefore, to ensure good
spectral quality and accurate line ratios, we have adopted an
H$\beta$ equivalent width limit of 8 \AA\ for the ensuing analysis. 

\begin{figure*}[htp]
\vskip -1.3in
\epsfxsize=5.5in
\hskip 1.0in
\epsffile{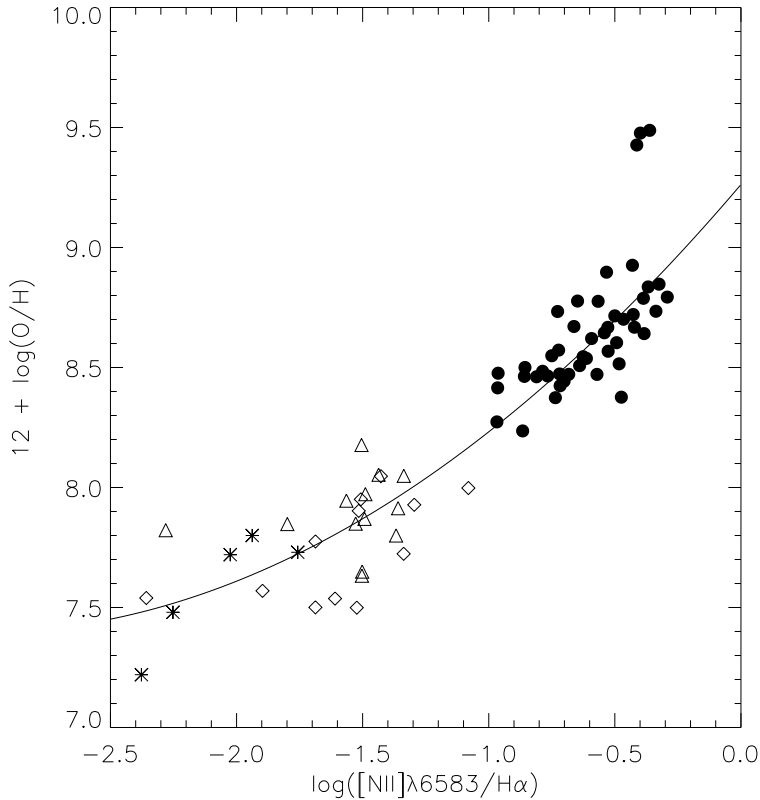}
\vskip -0.7in
 \figcaption[jmelbourne.fig4.ps]{This line diagnostic diagram relates the
  \nha line ratio to metallicity.
  Abundances calculated using the $R_{23}$ fit are shown
  as circles, while $p_3$ abundances are shown as triangles.
  Abundances calculated with the T$_e$
  method are shown as diamonds.  T$_e$ abundance data taken from Izotov et
  al. (1997) are shown as '*'s. We fit a quadratic function to the
  data (solid line, see 
  text).
  \label{fig:N2diag}}
\end{figure*}

With the equivalent width limit of H$\beta$ and the nitrogen line
ratio criteria specified, we calculate metallicities for 59 additional
galaxies, 46 using the $R_{23}$ method and  
13 using the $p_3$ method.  The metallicities for these galaxies along
with the oxygen line ratios are
given in Tables \ref{table:R23met} and \ref{table:P3met}.   We also have
$p_3$ abundances for the twelve Group I galaxies that have T$_e$ 
metallicities.  Seven of the galaxies have $p_3$ abundances within
0.1 dex of their T$_e$ result.  Four are within 0.2 dex and one galaxy
is highly deviant at 0.4 dex.  We believe the reason for the deviation
is related to the 
evolutionary state of the starburst. This particular galaxy appears to
be a highly evolved, low-metallicity starburst.  Since it is well past
the peak of its star formation episode, it has an
elevated [OII]$\lambda3726+\lambda3729$/H$\beta$ line ratio despite the
low overall metal abundance.  The Pilyugin fit may not track this
evolution as his fit is based on the relatively young starbursts
studied by Izotov et al. (1997; 1998; 1999).  Further comparison of
the two methods will be presented in Paper II.    
When calibrating the low metallicity end of our line diagnostic
diagrams, we will use the T$_e$ abundances when available and the
$p_3$ abundances for objects with no T$_e$ result.

\subsection{Metallicities from Line-Diagnostic Diagrams}    
We use the $R_{23}$, $p_3$ and T$_e$ metallicities presented in the
previous sections to relate the \ohb and \nha line ratios with metallicity.  
We choose these two line ratios because they are
observed in nearly all of our ELGs, typically have good
signal-to-noise ratios, and are fairly insensitive to uncertainties in
reddening 
corrections. We find that both  the \ohb and \nha line ratios can be used as
predictors of metallicity.  However, they both have limitations. 
A plot of the relationship between \OHB and \NHA is given in Figure
\ref{fig:O3N2}.  In this plot, metallicity varies smoothly over the
distribution of galaxies with low metallicity systems in the upper left and
high metallicity galaxies in the lower right.  At the low metallicity
end, the \ohb line remains almost constant for a large range of \NHA values.
Specifically, for \NHA $<$ -1.2 the \ohb line ratio is not a good
metallicity indicator.  Similarly, on the high metallicity end the \nha
line ratio ceases to be a good metallicity indicator for \OHB $<
-0.25$ (see also Figure 2).  We will use this information to decide
which line ratio 
provides the most accurate metal abundance for a given galaxy.

In Figure \ref{fig:N2diag} we plot 
\NHA vs. metallicity for the Group I and II data.  As the data are
sparse on the low metallicity end  we supplement with T$_e$ 
results  taken from the literature (Izotov et al. 1997).  The diagram
shows that metallicity  
increases with nitrogen line strength as a smooth single valued function up to 
the metal rich end of the distribution.    At \NHA $ > -0.45$ 
there is a sharp upturn in the observed distribution.  This 
corresponds  to the regime where the nitrogen line ratio ceases to be
a good
metallicity indicator, precisely the phenomenon seen in Figures 2 and
\ref{fig:O3N2}. We fit a quadratic function to the data points with
\NHA $< -0.45$ and obtain the following result:
\begin{eqnarray}
12 + \mbox{log(O/H)} & = & 9.26 + 1.23 N + 0.204 N^2, \label{eqn:n2met} \\
N & = & \mbox{log ([NII]}\lambda6583/\mbox{H}\alpha). \nonumber
\end{eqnarray}
The data points have a root-mean-square (RMS) scatter about the fit
of 0.156 dex.  We use Equation \ref{eqn:n2met} to estimate the  
metallicity for galaxies with \OHB $> -0.25$.    

\begin{figure*}[htp]
\vskip -1.3in
\epsfxsize=5.5in
\hskip 1.0in
\epsffile{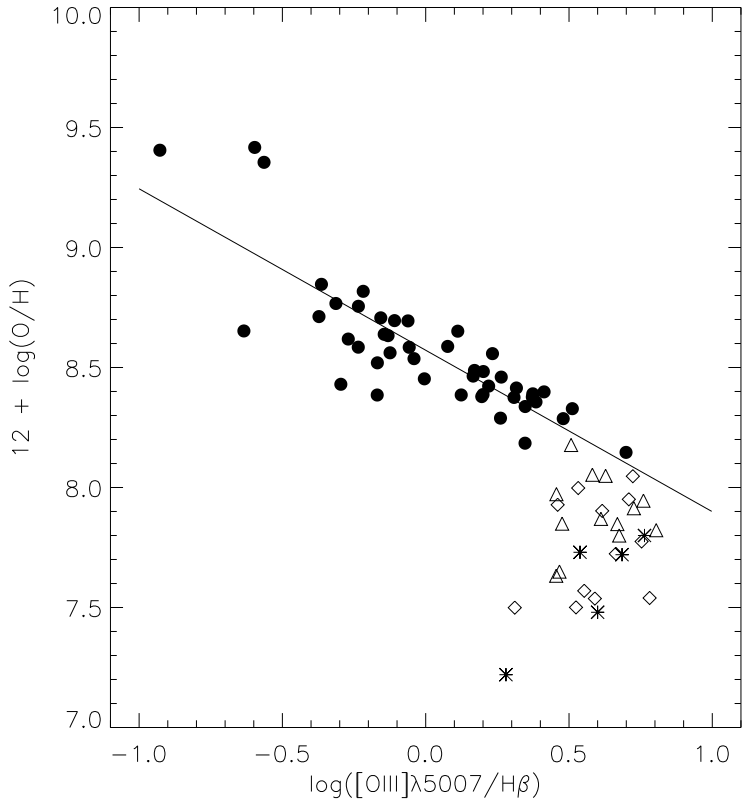}
\vskip -0.7in
\figcaption[jmelbourne.fig5.ps]{This line diagnostic diagram relates the
  \ohb line ratio to metallicity. Abundances calculated using the $R_{23}$ fit are shown
  as circles, while $p_3$ abundances are shown as triangles.
  Abundances calculated with the T$_e$
  method are shown as diamonds.  T$_e$ abundance data taken from Izotov et
  al. (1997) are shown as '*'s. We fit a line to the high
  metallicity branch (solid line, see text). 
  \label{fig:O3diag}}
\end{figure*}

In Figure \ref{fig:O3diag} we plot \OHB vs 
metallicity for the Group I and II data.  The relation is
well defined at the high metallicity end.  On the low metallicity end,
we see a scattered clump of galaxies with metallicities ranging from
7.5 to 8.2 dex and \OHB ranging from 0.3 to 0.8.  Again
we see that \ohb is not a good indicator of metallicity for low
metallicity systems as predicted by Figure \ref{fig:O3N2}. We fit a linear
function to all points on the upper metallicity branch, 
where \NHA $>$ -1.2.  We
ignore objects in the turn-around region and below.  The result
for the fit is as follows:
\begin{eqnarray}
\label{eqn:o3met1}
12 + \mbox{log(O/H)} & = & 8.65 - 0.663 O_x  \\
O_x &=& \mbox{log([OIII]}\lambda5007/\mbox{H}\beta), \nonumber
\end{eqnarray}
with an RMS of 0.149 dex. 
We use Equation \ref{eqn:o3met1} to estimate
metallicities for KISS galaxies with \NHA $>$ -1.2.

We use Equations \ref{eqn:n2met} and \ref{eqn:o3met1} to compute final
metallicity estimates for all KISS ELGs with the necessary spectral
information available.  Included are all objects that possess
follow-up spectra rated as good or excellent quality, that have been
classified as starbursting ELGs, and for which measurements of both
\nha and \ohb exist.  A total of 519 galaxies satisfy these criteria.
For a number of galaxies, metallicity estimates are computed using a
single line ratio, while for others both line ratios are used.  
We  combine the
nitrogen and oxygen metallicity results in the following way. 
Referring again to Figure \ref{fig:O3N2},
for galaxies with \NHA $<$ -1.2 we calculate the metallicity using
only the nitrogen
line ratio and Equation \ref{eqn:n2met}.   
For objects with  \OHB $<$ -0.25, we
use only the oxygen line ratio and  Equation
\ref{eqn:o3met1} to calculate metallicity.  For galaxies with \NHA $>$
-1.2 and \OHB $>$ -0.25 we calculate metallicities using both the
nitrogen and oxygen line ratios and  Equations
\ref{eqn:n2met} and \ref{eqn:o3met1}.  We take the average of the two
results to produce a final abundance estimate.  When we have
estimates of the metallicity from both the nitrogen and oxygen line
ratios, we find that the RMS scatter in the
difference of the two measurements is 0.13, but that the mean
difference is -0.009 (i.e., consistent with zero difference). 

This process generates metallicity estimates for 519 homogeneously observed
starbursting galaxies out to a redshift of z=0.095.  Since the
RMS scatter in Equations \ref{eqn:n2met} and \ref{eqn:o3met1} is 0.156
and 0.149 respectively, we assign a 
uncertainty of 0.16 dex to each metallicity measurement.  To be
conservative, we use this uncertainty even when averaging two
metallicity estimates together.   Abundance
estimates for individual galaxies calculated from these empirical methods
will be tabulated in Gronwall et al. (2002b).  Here we use the
results to identify low metallicity candidates for further study.  A list of
objects with metallicities below 7.9 is given in Table
\ref{table:lowmet}.  Several of these objects have been observed in
detail with the Lick telescope (Paper II) but
many more warrant high signal-to-noise ratio observations yielding T$_e$
abundance data.  As the objects in Table \ref{table:lowmet} are low
metallicity, high quality abundance measurements of these galaxies can be
used to study the primordial helium
abundance (Izotov et al. 1994; 1997) and place constraints on
Big Bang nucleosynthesis.  In the next section, this large
sample of data will be used to investigate the existence and form of the
metallicity-luminosity relation.   

\section{The Metallicity-Luminosity Relationship}
We combine galaxy metallicity estimates with 
calculations of the absolute B magnitudes to investigate 
the form of the metallicity-luminosity relationship.  Apparent magnitudes,
corrected for galactic reddening, are 
measured from the imaging portion of the survey data.  We adopt a Hubble
constant of H$_0 = 75 $ km$ \cdot$ s$^{-1}  
\cdot $Mpc$^{-1}$, and use the redshift measured from the follow-up
spectrum of each galaxy to arrive at the absolute magnitude. 

\begin{figure*}[htp]
\vskip -2.1in
\epsfxsize=6.5in
\hskip 0.5in
\epsffile{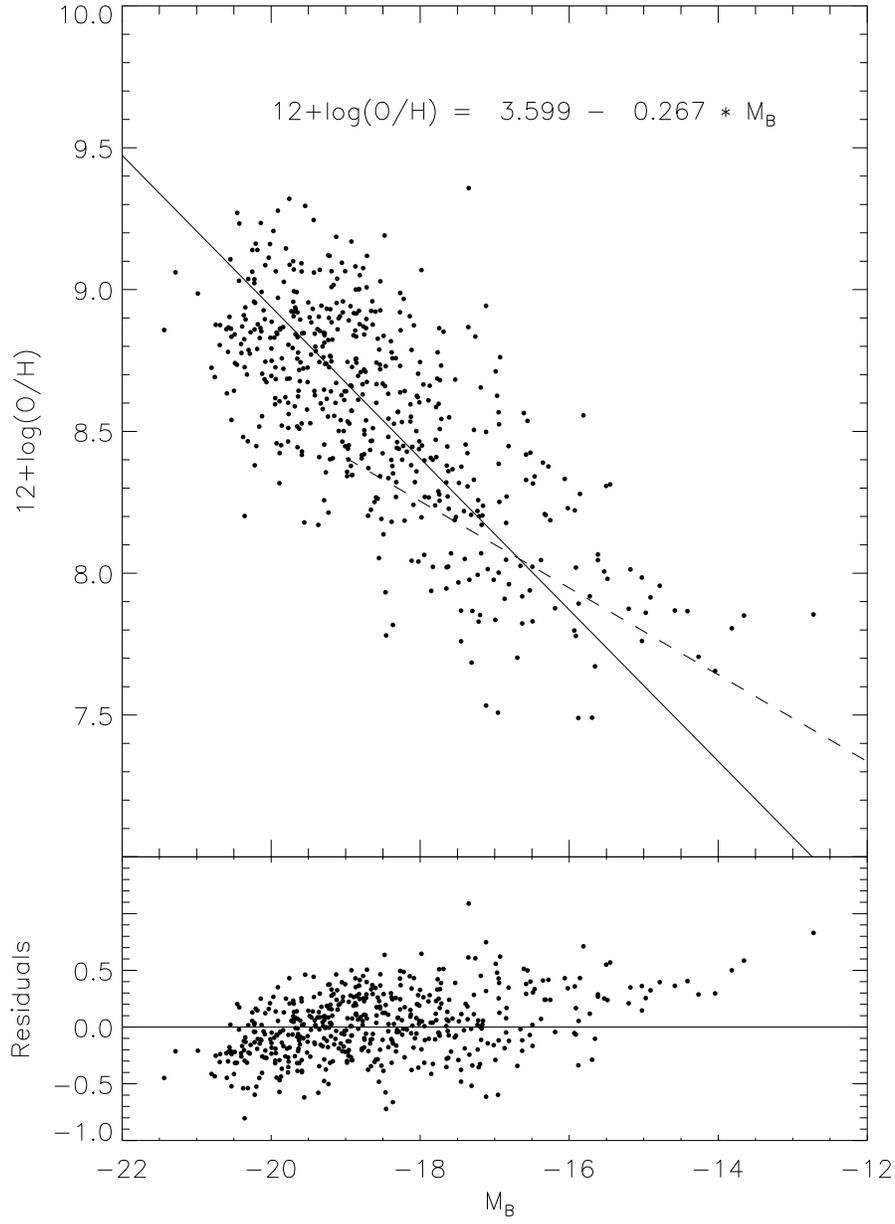}
\vskip -0.1in
\figcaption[jmelbourne.fig6.ps]{The metallicity-luminosity relationship is
  plotted in the top  
  panel.  The fit to the relationship is given as a solid line while
  the fit derived by Skillman et al. (1989) is shown as a dashed line.
  The bottom panel 
  shows the residuals to the fit which have an RMS deviation of
  0.27. 
  \label{fig:metlumFIN}}
\end{figure*}

Our metallicity-luminosity relation is shown in Figure
\ref{fig:metlumFIN}.  The general trend is an increase in metallicity
with luminosity over  
the full magnitude range of the data M$_B$ = -21 to -12.  Using a
bivariate linear least squares fitting 
technique,  we first fit the
data with the absolute magnitude as the independent variable, then
fit the data with the metallicity as the independent variable.  Our
final linear fit is the mean of the two fits, which is described by
the equation:
\begin{eqnarray}
\label{eqn:metlumpag}
12 + \mbox{log(O/H)} &=& 3.60  - 0.267 M_B.\\
& &\hspace{-.3in}(\pm 0.20) \hspace{.2in}(\pm0.009) \nonumber
\end{eqnarray}
This is shown as the solid line in Figure 6.
The scatter about the fit remains constant over the absolute
magnitude range of 
$M_B=$ -21 to -16, with an RMS of 0.27 dex.  The scatter is
systematically one sided for $M_B > -16$, implying that there may be a
shallower slope at the low metallicity end.  The small formal errors in
the coefficients of Equation \ref{eqn:metlumpag} are somewhat deceiving.
In arriving at these error bars we assumed an error in the absolute
magnitude of 0.5 mags, which reflects the formal photometric errors,
the uncertainty in the Hubble constant, and any possible peculiar velocities of
the individual galaxies.  An error in the
metallicity of 0.16 dex is assumed,  
consistent with the scatter in the line diagnostic diagrams from which
we estimate our metallicities.  Because we have a large data set, these
error estimates translate to very small formal errors in the
slope and intercept of the fits.  However, the difference in slopes from the
direct and inverse fits of the bivariate fitting are substantially
larger than the quoted uncertainty.

The slope we obtain is
affected by two parameters.   First, the choice of the $R_{23}$
calibration affects the metallicities we derive for our Group II and
therefore Group III data.  Second, the absolute magnitude
measurements may suffer from internal absorption.  More massive
galaxies are likely to have systematically higher extinction than 
dwarf galaxies due to their higher metallicities and higher dust
content.  This would, if left uncorrected, affect the slope of 
the metallicity-luminosity relation.  We attempt to address both these issues.

There are several different calibrations relating
$R_{23}$ to metallicity at the high metallicity end.  We chose the
Edmunds and Pagel 
(1984) fit primarily because of its long standing use and simplicity.  
Pilyugin (2000) offers an updated $R_{23}$ relation which takes into
account data not available when Edmunds and Pagel made their
original result.  The Pilyugin relation is:
\begin{equation}
12 + \mbox{log (O/H)} = 9.50 - 1.40 \cdot log(R_{23}),
\end{equation}
which is roughly parallel to but systematically lower
than the Edmunds and Pagel fit by 0.07 dex. If we use this result to
calibrate the 
upper branch of the $R_{23}$ relation, we find the following
metallicity-luminosity relation:
\begin{eqnarray}
12 + \mbox{log (O/H)} = 3.763 - 0.255 M_B.
\end{eqnarray}
The derived slope is slightly shallower than when using the Edmunds and Pagel
$R_{23}$ relation. However, it remains much steeper than previous
studies (see below).
We conclude that while our result is somewhat dependent on the
$R_{23}$ calibration, any reasonable choice for $R_{23}$ will give rise
to a significantly steeper slope in the metallicity-luminosity
relation, compared with the slopes derived for dwarf galaxies.

\begin{figure*}[htp]
\vskip -2.1in
\epsfxsize=6.5in
\hskip 0.5in
\epsffile{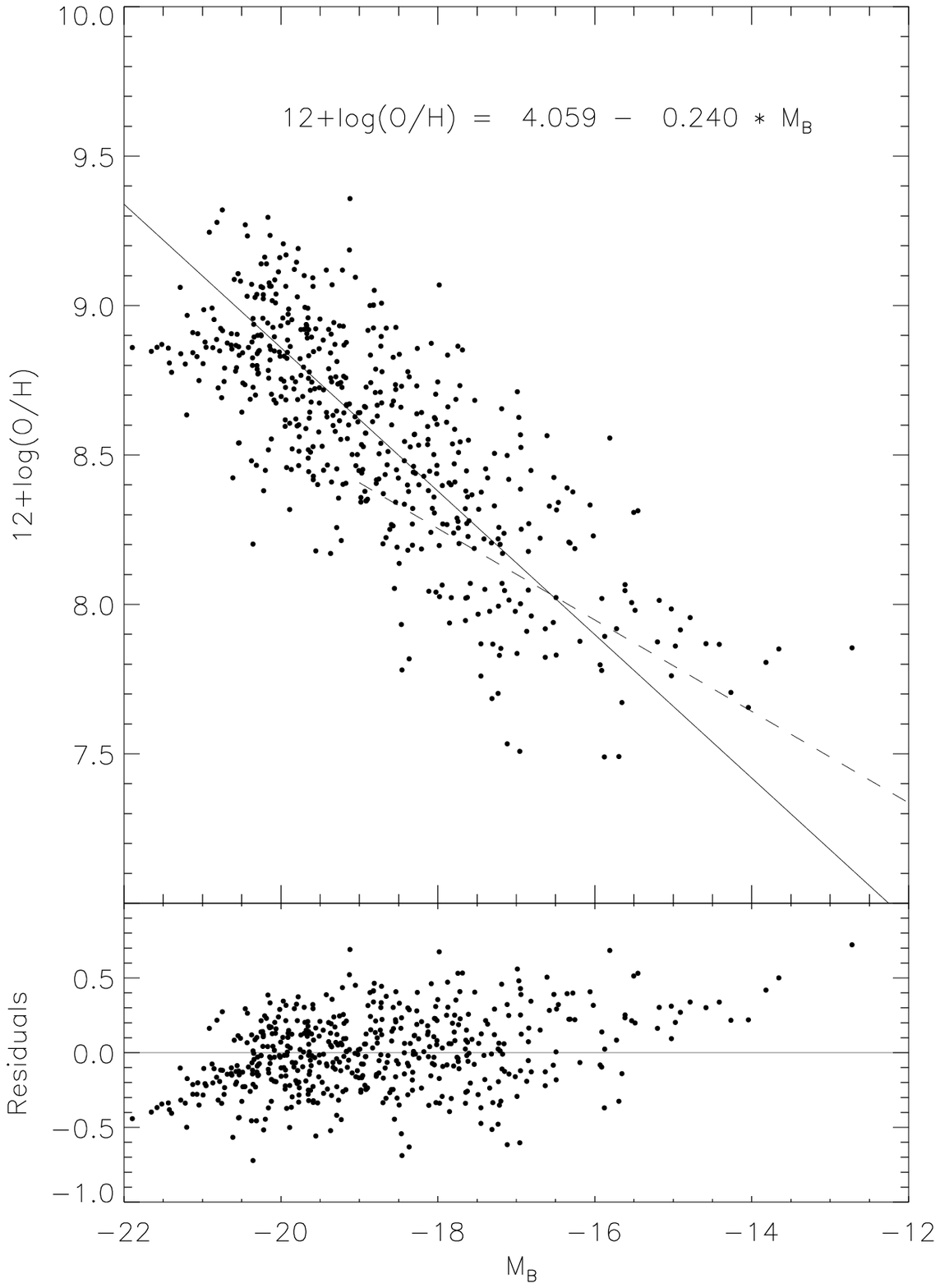}
\vskip -0.1in
\figcaption[jmelbourne.fig7.ps]{After correcting the luminosities for reddening
  and adopting the Pagel $R_{23}$ relation, we arrive at the
  metallicity-luminosity relationship plotted in the top  
  panel.  The fit to the relationship is given as a solid line while
  the fit derived by Skillman et al. (1989) is shown as a dashed line.
  The bottom panel 
  shows the residuals to the fit which have an RMS deviation of
  0.25. 
  \label{fig:metlumcor}}
\end{figure*}

More problematic is quantifying the absorption caused by dust
internal to a given galaxy.  As massive galaxies tend to
contain more dust, they suffer from internal absorption more 
than dwarf galaxies.  Thus luminosities of massive galaxies are likely
to be underestimated more than luminosities of dwarf galaxies.
Correcting the measured absolute magnitudes for this internal 
extinction will make the slope of the 
metallicity-luminosity relation more shallow.  As the imaging data do
not in general reveal the
Hubble type or axial ratio of the galaxies, traditional absorption
corrections are not possible.   We attempt a non-standard  
correction in the following way. 
A plot of B-V color vs. M$_B$ for each galaxy shows a trend of
increasing red colors with 
increasing  luminosity.  There is  scatter in the trend due both to
the variations in the stellar populations from galaxy to galaxy, as well as to
the amount of internal reddening and absorption in the galaxy.  We fit
a line to the trend ignoring the 
most deviant red points which are most strongly affected by dust.  The
fit is taken to be the center of the  
color distribution in the absence of internal reddening.  We then
infer a color excess, E(B-V), by measuring the color difference
between each galaxy and the regression line.  We assume
that objects redder than $2\sigma$ from the mean regression line, and  with
$c_{H\beta}$ greater than 0.25, suffer from internal absorption.  The
deviant points are corrected to  
the trend line by assuming the following reddening law:
\begin{equation}
A_B = 4.0 E(B-V) \nonumber
\end{equation}      
While admittedly ad hoc, this absorption correction does
systematically account for the most heavily extincted galaxies in our
sample. All of the galaxies that were corrected have M$_B <$ -16 mags.  The
result of the correction on the metallicity-luminosity relation (using
the Edmunds and Pagel $R_{23}$ relation) is 
plotted in Figure \ref{fig:metlumcor}.
Once again a bivariate linear least squares fit is applied and we find
the following result for the metallicity-luminosity relation:
\begin{eqnarray}
12 + \mbox{log (O/H)} = 4.059  - 0.240 M_B,\\
& &\hspace{-1.6in}(\pm 0.17) \hspace{.2in}(\pm0.006) \nonumber
\end{eqnarray}
with an RMS of 0.252.  

After the extinction correction is applied to the
data, we continue to find a slope steeper than those reported in
the literature. Previous groups have concentrated on
the low-metallicity, low-luminosity end where T$_e$ abundances are
available.  Skillman et al. (1989) used 
metallicities from 20 nearby irregular galaxies,  with absolute
magnitudes between  
M$_B$ = -19 to -10.5, to study the metallicity-luminosity relation.  They  
calculated  metallicities from the T$_e$ method and distances from Cepheid 
variables and group associations, and found \(12 + \mbox{log(O/H)} = 5.50 - 
0.153 M_B\) (dashed line in Figure \ref{fig:metlumFIN}), with an RMS
deviation of 0.16 dex.  Similarly, Richer and McCall  
(1995) report a metallicity-luminosity relation for 18 nearby dwarf irregulars 
of \(12 + \mbox{log(O/H)} = (5.67\pm0.48) - (0.147\pm0.029) M_B\).  
Their data range in luminosity from M$_B$ = -18 to -10.5 with a dispersion 
that increases for M$_B >$ -15.  Again, metallicities are 
derived from the T$_e$ method and distances from stellar calibrators.


Until recently, published fits of the metallicity-luminosity
relation have not included high luminosity galaxies in their samples.
However, in looking back to data sets on the metallicity of large
spirals such as Zaritsky et al. (1994) we find previous evidence for
the fact that massive
galaxies follow a steeper metallicity-luminosity relation than do
dwarf galaxies.  This is clear from Figure 13 of Zaritsky et al., 
where the data follow a steeper slope than the line included on
the plot which has a slope similar to Skillman et al.  When we combine
the Zaritsky et al. data set (39 spirals) with the Skillman et
al. data, we find a metallicity-luminosity relation given by $12 +
$log(O/H)$ = 4.71 - 0.210 M_B$.   
In another effort to combine the low luminosity results with high
luminosity systems, Pilyugin and Ferrini (2000) combine the Richer and
McCall data set with 13 objects from the Garnett et al. (1997) data set
and include 17 objects from their own observations.  These data,
ranging in M$_B$ from -21.5 to -10.5,  have a 
metallicity-luminosity relation with a slope of
-0.192.  This is evidence that the overall slope may be steeper than
indicated by the dwarfs alone, though still not as steep as our data
indicate.  One might well expect our data to give a steeper slope than
the compilation of Pilyugin and Ferrini.  The large spirals used in
their data set have metallicities  derived from several HII regions
within the disk of each galaxy and averaged together.  In contrast, our
most metal rich galaxies tend to be starburst nucleus galaxies,
where the emission is coming from a large star-forming event in the
center of the galaxy.  Because spirals are known to exhibit radial
abundance gradients, nuclear starburst galaxies should as a group tend
to have higher metallicities than those measured from disk HII regions,
even for galaxies with similar luminosities.  This is born out
in a recent paper by 
Contini et al (2001).  They plot a metallicity-luminosity relation for
a large sample of galaxies.  The irregular galaxies in their sample follow a
metallicity-luminosity relation similar to Richer and McCall (1995).
The larger HII galaxies and UV selected galaxies follow a linear relation
with a higher slope of -0.173.  Displayed on their plot, but not included in
either fit, is a large population of starburst nucleus galaxies.
These galaxies tend to have luminosities similar to the UV selected
galaxies but tend to have higher metallicities.  It is clear that the
inclusion of the starburst nucleus galaxies in a composite fit would
lead to a significantly higher slope.  

In conclusion, we believe that the steeper slope indicated by our
data is reasonable.  We
stress, however, that this result is not meant to replace the work
of Skillman et al. and others, but rather to investigate the
metallicity-luminosity 
relationship with a wider sample of galaxy types.  In fact Figure 7
shows evidence that the overall form of the relation may not be a
simple linear function, but rather may be a
higher order polynomial.  The slope appears to become more shallow
at the low luminosity end, resembling the slopes found by previous
studies.  Because 
of the large scatter and relative paucity of galaxies at the low
luminosity end, we do not feel we can adequately justify a higher-order
fit with the current data set.   

\section{Discussion and Conclusions}

The metallicity-luminosity relation has significant implications for
galaxian evolution.  It appears to be  a continuous smooth
function from high-luminosity massive spirals to the low-luminosity
dwarf galaxies, implying that the metallicity-luminosity relation is at
work on all mass scales and all galaxy types with significant star
formation. The form of the metallicity-luminosity relation, along with
its intrinsic scatter, should
provide a useful constraint for theoretical models of chemical
evolution in galaxies. 

We have confirmed the existence of the
metallicity-luminosity  
relationship using data from 519 starburst galaxies.  We found a
linear relation between metallicity and absolute B magnitude given by:
\(12 + \mbox{log(O/H)}  = 4.059  - 0.240 M_B.\) 
Metallicities were 
derived for the large sample from secondary metallicity indicators
relating the strong  nebular lines, [OIII]$\lambda$5007 and 
[NII]$\lambda$6583, to metallicity.  We
used T$_e$ abundances from 12 galaxies, $p_3$ abundances for 13
galaxies and $R_{23}$
abundances for 46 galaxies, to relate emission-line ratios with
metallicity.  This study uses the largest sample of galaxies to date
to construct a metallicity-luminosity relation.  We find a
significantly steeper slope than previous results, most likely due to
the fact that we probe to higher luminosities and include a  more diverse
mix of galaxy types in our sample.  The large scatter, 0.252 dex,  
remains roughly constant to an absolute magnitude of $M_B =$ -16,
at which point the scatter becomes one sided.  This indicates that 
the slope may be  more shallow at the low luminosity
end, in agreement with previous results that have
focused on dwarf systems.  Interestingly, Richer and McCall have also found
the scatter 
in their metallicity-luminosity relation to be one sided for
objects with M$_B >$ -16.  This may be further evidence for the need
of higher order fits to the relation.  We have resisted the urge to
carry out a higher order fit to the data due to the large intrinsic
scatter in our sample. 

In interpreting the results illustrated in Figure 7, the reader 
should keep in mind the following cautionary note.  As mentioned above,
the abundances we measure for the more luminous KISS galaxies are 
predominantly central/nuclear values, rather than disk HII regions.
Since spiral galaxies typically exhibit radial abundance gradients, our 
abundances may be biased to higher values compared to previous studies.  
Clearly, one needs to consider the location within the galaxy
when quoting abundances (particularly for large spirals) and investigating
the metallicity-luminosity relation.  While there is no absolute correct
method or convention, we would argue that our use of nuclear abundances
is no less appropriate than using outlying disk HII regions (which will
be biased to lower abundances).  In fact, since dwarf galaxies tend
to exhibit no substantial abundance gradients, the use of central
abundance measurements might well be preferred.

We have investigated whether this effect could be the cause for our 
steeper metallicity-luminosity relation by comparing our results with 
those from previous studies.  For example, the large dataset of Zaritsky
et al. (1994), derived from disk HII regions rather than nuclear star-forming
regions, exhibits a metallicity-luminosity relation very similar to ours
at M$_B$ $<$ $-$18.  No bias is evident when the two datasets are compared
directly.  This may be due in part to the fact that many spiral galaxies 
exhibit only very shallow abundance gradients (or in some cases none at 
all).  We conclude that our use of central abundances is not the main
cause for our steep metallicity-luminosity relation, but rather that this
is a real phenomenon.  We point out that evidence for  a steeper slope
to the metallicity-luminosity relation has been available but not 
pursued until Pilyugin and Ferrini (2000) combined the Garnett et al. and
Richer and McCall data sets to observe the overall trend.  Their study
found a slope steeper than those indicated by Skillman et al. and
Richer and McCall.  Seeking further confirmation of this result, we
combined data from Zaritsky et al. (39 spirals) with data from
Skillman et al.  We again found evidence for a steeper slope.
The metallicity-luminosity 
relation using this data set is given by: $12 + $log(O/H)$ = 4.71 -
0.210 M_B$.  

The KISS sample of galaxies 
includes a diverse mass and morphology range, from blue compact 
dwarfs to giant nuclear starburst galaxies. By
including the more massive galaxies and achieving a 
large sample, we appear to be observing a more general
metallicity-luminosity trend than  is indicated by the dwarf galaxies
alone.   While the 
metallicity-luminosity relation derived by
Skillman et al. appears to be a good approximation of the low
metallicity data, extrapolating their relation to higher luminosities
is probably inappropriate.  Efforts to understand the physical
mechanisms that lead to the observed metallicity-luminosity relation
will be aided by the constraints set 
by the data presented in this paper.  

The methods used here to calculate coarse metal abundances
allow for a way to quickly identify low metallicity
candidates for further study.  The KISS galaxies believed to be low
metallicity will be targeted as high priority systems for abundance
quality spectra in future observing runs.  Eventually this will
yield a large number of low metallicity objects that can be used to
place constraints on the primordial helium abundance of the Universe.
The coarse abundance methods can also be applied to galaxies at higher
redshift, where 
[OIII]$\lambda$4363 is difficult to observe.  Metallicity comparisons
between galaxies over a range of redshifts will help to shed light on the
chemical evolution of galaxies.

\acknowledgements

We gratefully acknowledge financial support for the KISS project from 
an NSF Presidential Faculty Award to JJS (NSF-AST-9553020), as well as 
continued support for our ongoing follow-up spectroscopy campaign 
(NSF-AST-0071114).  We also thank Wesleyan University for providing
additional funding for several of the observing runs during which the
spectral data were obtained.  We are grateful to the anonymous referee
for many suggestions which improved the quality of this paper.  We
thank the numerous KISS team members 
who have participated in the spectroscopic follow-up observations during
the past several years, particularly Caryl Gronwall, Drew Phillips, Gary
Wegner, Laura Chomiuk, Kerrie McKinstry, Robin Ciardullo, Jeffrey Van Duyne
and Vicki Sarajedini.
Finally, we wish to thank the support staffs of Kitt Peak National Observatory,
Lick Observatory, the Hobbey-Eberly Telescope,  MDM Observatory, and Apache
Point
Observatory for their excellent assistance in obtaining these
observations.

\renewcommand{\arraystretch}{.6} 
\begin{deluxetable}{cccc}
\tablewidth{5 in}
\tablenum{1}
\tablecaption{Metal abundances from the $T_e$ method. \label{table:Te}}
\tablehead{\colhead{KISSR \tablenotemark{a}} &
  \colhead{log([OII]/H$\beta$)} & 
  \colhead{log([OIII]/H$\beta$) \tablenotemark{b}} & \colhead{12 + log(O/H)}}
\startdata
   49 &  0.619 &  0.670  & 8.00\\
   85 &  0.024 &  0.658  & 7.50\\
  396 &  0.413 &  0.759  & 7.90\\
  666 & -0.310 &  0.913  & 7.54\\
  675 &  0.145 &  0.878  & 7.78\\
 1013 &  0.438 &  0.829  & 7.72\\
 1194 &  0.311 &  0.834  & 7.95\\
 1490 &  0.179 &  0.715  & 7.54\\
 1752 &  0.038 &  0.678  & 7.56\\
 1778 &  0.466 &  0.599  & 7.93\\
 1845 &  0.393 &  0.847  & 8.04\\
\tableline \\
KISSB \tablenotemark{c}  & log([OII]/H$\beta$) &
log([OIII]/H$\beta$)$^b$ & 12 + log(O/H)\\
\tableline \\
   23 &  0.430 &  0.448  & 7.50\\
\tableline 
\enddata
\tablenotetext{a}{This is the running number for the galaxy as listed
in Salzer et al. 2001.}
\tablenotetext{b}{[OIII]$\lambda4959+\lambda5007$/H$\beta$}
\tablenotetext{c}{This is the running number for the galaxy as listed
  in Salzer et al. 2002a.}
\end{deluxetable}

\begin{deluxetable}{cccc}
\tablenum{2}
\tablewidth{5 in}
\tablecaption{Metal abundances from the $R_{23}$ method. \label{table:R23met}}
\tablehead{\colhead{KISSR } &
  \colhead{log([OII]/H$\beta$)} & 
  \colhead{log([OIII]/H$\beta$) \tablenotemark{a}} & \colhead{12 + log(O/H)}}
\startdata
                1 &  0.651 &  0.249  & 8.47\\
                3 &  0.729 & -0.044  & 8.47\\
               68 &  0.426 &  0.538  & 8.48\\
               91 &  0.549 &  0.291  & 8.55\\
              109 &  0.506 &  0.326  & 8.57\\
              110 &  0.434 &  0.016  & 8.78\\
              117 &  0.484 &  0.499  & 8.48\\
              124 &  0.506 &  0.497  & 8.46\\
              145 &  0.420 & -0.110  & 8.84\\
              185 &  0.355 & -0.093  & 8.90\\
              257 &  0.562 &  0.433  & 8.46\\
              265 &  0.489 &  0.389  & 8.54\\
              292 &  0.522 &  0.605  & 8.37\\
              317 &  0.404 &  0.637  & 8.42\\
              322 & -0.043 & -0.439  & 9.43\\
              326 &  0.631 &  0.120  & 8.54\\
              327 &  0.580 &  0.472  & 8.42\\
              333 &  0.430 & -0.188  & 8.85\\
              335 &  0.386 &  0.358  & 8.64\\
              343 &  0.415 &  0.824  & 8.24\\
              368 &  0.711 & -0.171  & 8.52\\
              398 &  0.501 & -0.020  & 8.72\\
              400 &  0.371 & -0.239  & 8.93\\
              515 &  0.524 &  0.067  & 8.67\\
              520 &  0.440 & -0.032  & 8.79\\
              577 &  0.572 & -0.110  & 8.67\\
              582 &  0.759 &  0.472  & 8.27\\
              595 &  0.005 & -0.803  & 9.48\\
              610 &  0.461 &  0.202  & 8.67\\
              643 &  0.505 &  0.441  & 8.50\\
              830 &  0.690 &  0.386  & 8.38\\
              833 &  0.566 &  0.001  & 8.64\\
              894 &  0.614 & -0.044  & 8.60\\
              910 &  0.416 &  0.063  & 8.78\\
             1016 &  0.502 & -0.006  & 8.72\\
             1028 & -0.092 & -0.472  & 9.49\\
             1032 &  0.520 &  0.295  & 8.57\\
             1055 &  0.550 & -0.145  & 8.70\\
             1062 &  0.565 &  0.345  & 8.51\\
             1402 &  0.571 & -0.509  & 8.73\\
             1416 &  0.615 &  0.325  & 8.47\\
             1424 &  0.490 & -0.247  & 8.79\\
             1537 &  0.522 &  0.510  & 8.44\\
             1885 &  0.364 &  0.237  & 8.73\\
             1940 &  0.626 &  0.321  & 8.46\\
             2021 &  0.564 &  0.084  & 8.62\\
\enddata
\tablenotetext{a}{[OIII]$\lambda4959+\lambda5007$/H$\beta$}
\end{deluxetable}

\pagestyle{empty}
\begin{deluxetable}{cccc}
\tablenum{3}
\tablewidth{5 in}
\tablecaption{Metal abundances from the $p_{3}$ method. \label{table:P3met}}
\tablehead{\colhead{KISSR } &
  \colhead{log([OII]/H$\beta$)} & 
  \colhead{log([OIII]/H$\beta$) \tablenotemark{a}} & \colhead{12 + log(O/H)}}
\startdata
   96 &  0.280 &  0.851  & 7.91\\
   97 &  0.493 &  0.753  & 8.05\\
  105 &  0.166 &  0.592  & 7.65\\
  120 &  0.167 &  0.800  & 7.80\\
  272 &  0.359 &  0.601  & 7.85\\
  310 & -0.097 &  0.929  & 7.82\\
  311 &  0.248 &  0.794  & 7.85\\
  404 &  0.459 &  0.581  & 7.97\\
  405 &  0.154 &  0.580  & 7.63\\
  528 &  0.507 &  0.707  & 8.05\\
  785 &  0.287 &  0.885  & 7.95\\
  803 &  0.600 &  0.632  & 8.18\\
 1794 &  0.322 &  0.736  & 7.87\\
\enddata
\tablenotetext{a}{[OIII]$\lambda4959+\lambda5007$/H$\beta$}
\end{deluxetable}

\pagestyle{empty}
\begin{deluxetable}{cccc}
\tablenum{4}
\tablewidth{5 in}
\tablecaption{Galaxies with estimated metallicities below
  7.9 dex. \label{table:lowmet}} 
\tablehead{\colhead{KISSR } & \colhead{
    log([OIII]/H$\beta$)\tablenotemark{a}} &
  \colhead{log([NII]/H$\alpha$)\tablenotemark{b}} & \colhead{12 + log(O/H)}}
\startdata
        55  &  0.808 &  -1.654 &  7.78 \\
        73  &  0.623 &  -1.531 &  7.85 \\
        85  &  0.533 &  -1.686 &  7.76 \\
       105  &  0.467 &  -1.502 &  7.87 \\
       272  &  0.476 &  -1.528 &  7.85 \\
       310  &  0.804 &  -2.281 &  7.51 \\
       311  &  0.669 &  -1.799 &  7.70 \\
       396  &  0.634 &  -1.515 &  7.86 \\
       404  &  0.456 &  -1.489 &  7.88 \\
       405  &  0.455 &  -1.503 &  7.87 \\
       471  &  0.676 &  -2.340 &  7.49 \\
       666  &  0.788 &  -2.345 &  7.49 \\
       675  &  0.753 &  -1.688 &  7.76 \\
       698  &  0.844 &  -1.577 &  7.82 \\
       785  &  0.760 &  -1.565 &  7.83 \\
       799  &  0.690 &  -1.566 &  7.83 \\
       803  &  0.507 &  -1.505 &  7.87 \\
       885  &  0.727 &  -1.834 &  7.68 \\
       986  &  0.619 &  -2.202 &  7.53 \\
      1194  &  0.709 &  -1.506 &  7.87 \\
      1490  &  0.590 &  -1.606 &  7.81 \\
      1752  &  0.553 &  -1.896 &  7.66 \\
      1794  &  0.612 &  -1.492 &  7.87 \\
\tableline \\
KISSB   & log([OIII]/H$\beta$)\tablenotemark{a} &
log([NII]/H$\alpha$)\tablenotemark{b} & 12 + log(O/H)\\
\tableline \\
      15 &  0.607 &  -1.556 &  7.84 \\
      23 &  0.323 &  -1.525 &  7.85 \\
      35 &  0.873 &  -1.651 &  7.78 \\
      41 &  0.786 &  -1.620 &  7.80 \\
      47 &  0.710 &  -1.792 &  7.71 \\
      53 &  0.660 &  -1.861 &  7.67 \\
      61 &  0.656 &  -1.586 &  7.82 \\
      66 &  0.770 &  -1.463 &  7.89 \\
\enddata
\tablenotetext{a}{[OIII]$\lambda$5007/H$\beta$}
\tablenotetext{b}{[NII]$\lambda$6583/H$\alpha$}
\end{deluxetable}

\end{document}